\RequirePackage{fix-cm}
\documentclass[twocolumn,epjc3]{svjour3}  
\smartqed  
\RequirePackage{graphicx}
\RequirePackage{xcolor}
\journalname{Eur. Phys. J. C}
\begin{document}
\sloppy

\title{Multiplicity Dependence of J/$\psi$ Production and QCD Dynamics in $p+p$ Collisions at $\sqrt{s}$ = 13 TeV
}


\author{Suman Deb \thanksref{addr1}
        \and
        Dhananjaya Thakur\thanksref{addr1} \and Sudipan De\thanksref{e2,addr1} \and Raghunath Sahoo\thanksref{e1,addr1}
}

\thankstext{e1}{Corresponding author: Raghunath.Sahoo@cern.ch}
\thankstext{e2}{Present Address: School of Physical Sciences, National Institute of Science Education and Research, HBNI, Jatni - 752050, India}

\institute{Discipline of Physics, School of Basic Sciences, Indian Institute of Technology Indore, Simrol, Indore 453552, INDIA \label{addr1}
}

\date{Received: date / Accepted: date}

\maketitle

\begin{abstract}
In inelastic $p+p$ collisions, the interacting objects are quarks and gluons (partons). It is believed that there are multiple interactions between the partons in a single $p+p$ event. Recent studies of multiplicity dependence of particle production in $p+p$ collisions have gathered considerable interest in the scientific community. According to several theoretical calculations, multiple gluon participation in hadronic collisions is the cause of high-multiplicity events. If the interaction is hard enough (large $p_{\rm T}$ transfer), the semi-hard processes of multiple interactions of partons might also lead to production of heavy particles like J/$\psi$. At the LHC, an approximately linear increase of the relative J/$\psi$ yield with charged particle multiplicity is observed in $p+p$ collisions. In the present work, we have studied the contribution of quarks and gluons to the multiplicity dependence of J/$\psi$ production using pQCD inspired event generator, PYTHIA8 tune 4C, in $p+p$ collisions at $\sqrt{s} =$13 TeV by investigating relative J/$\psi$ yield and relative $\langle  p_{\rm T} \rangle$ of J/$\psi$ as a function of charged particle multiplicity for different hard-QCD processes. We have estimated a newly defined ratio, $r_{pp} = {\langle  p_{\rm T}^{2} \rangle}_{i}/{\langle  p_{\rm T}^{2} \rangle}_{\rm MB}$, to understand J/$\psi$ production in 
high-multiplicity $p+p$ collisions. For the first time we attempt to study the nuclear modification factor like observables ($R_{\rm pp}$ and $R_{\rm cp}$) to understand the QCD medium formed in high-multiplicity $p+p$ collisions.
\end{abstract}

\section{Introduction}  
\label{intro}
Development of novel theoretical models to understand quarkonia production mechanism in hadronic collisions is one of the challenging tasks in high-energy particle/nuclear physics. A comprehensive review of the status of understanding of quarkonia  in the present time and specific recommendations for further progress is reported in Ref.~\cite{Brambilla:2010cs}. Several theoretical models like, Color Singlet, non-relativistic QCD (NRQCD) and the Color Evaporation Model try to explain the charmonia production
in hard processes~\cite{Andronic:2015wma,Vogt:2019zmr,Ma:2016exq}. There have been dedicated efforts~\cite{Andronic:2015wma,Butenschoen1,Butenschoen:2012px,Cheung:2018tvq,Cheung:2017osx,Zhao:2007hh,Grandchamp:2001pf}  in understanding the production cross section and polarisation of J/$\psi$ by taking the inputs from recent LHC measurements~\cite{Aad:2014rua,Aaboud:2016fzt,Acharya:2018uww,BatistaCamejo:2017zvu}. At the LHC, the event multiplicity dependence of charmonium production has garnered a considerable interest among the scientific community which is thought to provide the event structure in $p+p$ collisions in terms of interplay between soft and hard interactions. 

ALICE experiment has measured the relative J/$\psi$ yield, $(dN_{\rm J/\psi}/dy)/(\langle  dN_{\rm J/\psi}/dy \rangle)$, as a function of relative charged particle multiplicity density, $(dN_{\rm ch}/dy)/(\langle  dN_{\rm ch}/dy \rangle)$ in $p+p$ collision at $\sqrt{s}$ = 7 TeV in both mid-rapidity ($|y|<0.9$) and forward rapidity (2.5 $< y <$ 4.0)~\cite{Abelev:2012rz}. It is observed that the relative J/$\psi$ yields are consistent with a linear, or stronger than linear increase as compared to relative charged particle multiplicity density at forward and mid-rapidity~\cite{Weber:2017hhm,Thakur:2019qau}. There have been several theoretical models~\cite{Srivastava:2017bcm,Weber:2018ddv,Ferreiro:2012fb,Kopeliovich:2013yfa} to explain this behaviour. In this regard, one of the well-known Monte Carlo event generators PYTHIA8~\cite{pythia8html}, which includes multiple-parton interactions (MPIs), rescattering~\cite{Corke:2009tk} and color reconnections (CR)~\cite{Cuautle:2016ukm,Thakur:2017kpv} can be used to have an insight into the observed behaviour.
 These mechanisms together lead to a more realistic simulation of high-energy $p+p$ collisions. 
 
 MPIs are multiple and independent $2\rightarrow2$ scatterings. They generally contribute to the underlying events. However, in high-energy collisions, more than one collision could be hard enough to produce a J/$\psi$ leading to an increase in the J/$\psi$ multiplicity with the number of charged particles produced in an event. PYTHIA8 has added rescattering-- a natural consequence of MPIs. This is because, a parton from one proton could interact with two independent partons from the other proton (essentially a $2\rightarrow2$ scattering) followed by rescattering of one of the final-state partons from the hard scattering resulting in substantial increase in rescatterings. In addition, color reconnections between partons are also taken into account. MPIs can lead to overlapping color strings. Reconnection reduces string length, essentially giving low $p_{T}$ systems an effectively larger spatial extent in the transverse direction. Both rescattering and color reconnection can lead to J/$\psi$ production in sufficiently hard interactions. In other words, out of {\it n}-independent parton scatterings in a given p + p collision, some subset of which, $n'$, are hard collisions that could lead to J/$\psi$ production in the final state. Thus the J/$\psi$ can be produced in initial hard scatterings, if the final state is selected in the PYTHIA input deck, or as one of many possibilities in ``all QCD processes", which include all interactions arising from MPIs. 
 
The production of heavy quark is implemented in PYTHIA as perturbative scattering processes: gluon fusion ($gg \rightarrow Q\bar{Q}$, Q is charm or bottom) or light quark-antiquark annihilation ($q\bar{q} \rightarrow Q\bar{Q}$). Also, heavy quarks can be present in the parton distribution function leading to heavy quark production via $Qg \rightarrow Qg$ and $g \rightarrow Q\bar{Q}$. For the production of quarkonia, the perturbative scattering processes lead to NRQCD channels via colour-singlet and colour-octet pre-resonant states as included in PYTHIA. For colour-octet states, one additional gluon is emitted in the transition to the physical colour-singlet state. During the hadronization, a heavy quark is connected to a corresponding heavy antiquark, if they are close in phase space and bind to produce a quarkonium bound state. In Ref.\cite{Weber:2018ddv}, it has been demonstrated how different production mechanisms contribute to J$/\psi$ production in $p+p$ collisions and how they behave with MPIs.
 The production of J/$\psi$ can be explicit and as a part of the MPI framework {\it e.g.} the process like $gg  \rightarrow$ J/$\psi$ along with $gg \rightarrow q \bar{q} \rightarrow$ J/$\psi$ + X in the same $p + p$ interaction, where $X$ denotes multiplicity of particles in the final state. In the present work, we study the leading order (LO) contribution of $2 \rightarrow 2$ processes such as $gg \rightarrow C$ and $q\bar{q} \rightarrow C$ (where $C$ is defined as final state charmonium) to  J/$\psi$ events in hadronic collisions as a function of multiplicity \cite{Gluck:1977zm}. We have compared these processes with J/$\psi$ from MPIs which include ``all Hard-QCD processes"\footnote {\textbf{J/$\psi$ from all Hard-QCD processes:} \
 \small {It includes all the NRQCD processes (LO processes) such as gluon fusion, quark annihilation, flavor excitation and the semi-hard MPIs, which itself is governed by NRQCD. For colour-octet states, one additional gluon is emitted in the transition to the physical colour-singlet state. Production of any $^3S_1$, $^3P_J$ and $^3D_J$  states via colour-singlet and colour-octet mechanisms are included through Onia process in PYTHIA8. In order to have quarkonia (here charmonia is of our interest) production,~\textit{ Charmonium:all} flag in PYTHA8 is included, which allows quarkonia production through NRQCD framework~\cite{Shao:2012iz,Caswell:1985ui,Bodwin:1994jh}. The production of all the states of charmonia are included in PYTHIA8 and their decay products have significant contribution to J/$\psi$. The heavier charmonium states can decay into J/$\psi$ meson by emitting photons or pions which contribute to the inclusive J/$\psi$ production.}}.

In the second part of the paper, we have studied the softening and hardening of the $p_{\rm T}$-spectra of J/$\psi$ in different multiplicity classes with respect to minimum biased (MB) events using PYTHIA8. So far this type of study has been done only for light-flavor particles in experiments \cite{Andrei:2014vaa}, because of limited statistics for heavy-flavor particles. The present study will help to understand the difference in production mechanisms of J/$\psi$ compared to light-flavor particles. According to MPI, J/$\psi$ can be produced from first $2\rightarrow2$ hard scatterings as well as subsequent secondary hard scatterings~\cite{PorteboeufHoussais:2012gn}.  The trend of ${\langle  p_{\rm T} \rangle}_{J/\psi}$ as a function of charged particle multiplicity can shed light on hard and semi-hard contributions of MPI to J/$\psi$ production. In the high-multiplicity $p+p$ collisions, many interesting results have been found~\cite{ALICE:2017jyt,Alver:2010ck,Khachatryan:2010gv}, where a possible formation of a deconfined medium~\cite{Shuryak:1980tp,Adams:2005dq,Mangano:2017plv} is under discussion. For example, the charged particle multiplicity measured in high-multiplicity $p+p$ collisions at $\sqrt{s} = 7$ TeV exceeds the charged particle multiplicity for peripheral Cu+Cu collisions at $\sqrt{s_{NN}}$ = 200 GeV \cite{Alver:2010ck}. Therefore, $p+p$ collisions at $\sqrt{s}$ = 13 TeV might be more interesting to look for heavy-ion-like observables and to have a direct comparison. In Ref.~\cite{Zhou:2014kka}, a new variable, $r_{\rm AA} = ({\langle  p_{\rm T}^{2} \rangle}_{\rm AA})/({\langle  p_{\rm T}^{2} \rangle}_{\rm pp})$ has been proposed to understand the J/$\psi$ production in heavy-ion collisions in SPS, RHIC and LHC energies.  In this paper, we have studied similar kind of observables in different multiplicity classes in $p+p$ collisions to study the similarities/differences between the systems created in heavy-ion and high-multiplicity $p+p$ collisions. We have used the modified formula for $p+p$ collisions as $r_{\rm pp} = ({\langle  p_{\rm T}^{2} \rangle}_{i})/({\langle  p_{\rm T}^{2} \rangle}_{\rm MB})$, where $\langle p_{\rm T}^{2} \rangle_{i}$ and $\langle p_{\rm T}^{2} \rangle_{\rm MB}$ are the averaged transverse momentum square of J/$\psi$ in $i^{th}$ multiplicity class and MB events, respectively. Another most important observable for studying the system created in heavy-ion collisions is the nuclear modification factor ($R_{\rm AA}$). This is defined as the ratio of invariant yield in A+A collisions with respect to invariant yield in $p+p$ collisions multiplied by average number of collisions, $\langle  N_{\rm coll} \rangle$. $R_{\rm AA}  \neq$ 1 indicates that there is a modification from nuclear effects, either due to the deconfined medium or due to cold nuclear matter effects. Here, we introduce a similar ratio in $p+p$ collisions following the basic concept to understand the QCD medium formed in high-multiplicity $p+p$ collisions. These studies can give an idea about possible biasing to the observables under study.  

The paper is organised as follows. Section~\ref{eventgen} presents event generation and analysis methodology. Results are discussed in Section~\ref{result}, which is divided into four subsections: multiplicity dependence of J/$\psi$ production in hard-QCD processes, transverse momentum and multiplicity dependence of J/$\psi$ production, multiplicity dependence of $r_{\rm pp}$ and nuclear modification factor-like ratios. Finally in Section~\ref{sum}, we summarize our work highlighting major observations and outlook. 
\\\\\\
\section{Event generation and analysis methodology}
\label{eventgen}
A detailed explanation on PYTHIA8 physics processes can be found in Ref. \cite{pythia8html}. In this paper, we have used 4C tuned PYTHIA8 \cite{Corke:2010yf} (Tune:pp = 5, in this scheme, modified multi-partonic interaction parameters give a higher and more rapidly increasing charged pseudorapidity plateau, for better agreement with some early key LHC numbers). We have included the MPI-based scheme of Colour Reconnection ($ColourReconnection:reconnect = on$) of PYTHIA8. In this study, we have simulated inelastic,~non-diffractive component of the total cross section for all hard QCD processes ($HardQCD:all=on$), which includes the production of heavy quarks. Beside these processes, we have also simulated leading order processes for heavy quark production, namely, $gg \rightarrow C$, $q\bar{q} \rightarrow C$ ($HardQCD:gg2ccbar=on$ and $HardQCD:qqbar2ccbar=on$), separately. A cut on transverse momentum, $p_{\rm T}$ $ \textgreater$ 0.5 GeV/c (using $PhaseSpace:pTHatMinDiverge$) is used to avoid the divergences of QCD processes in the limit $p_{\rm T}$ $\rightarrow$ 0. The number of multiple-parton interactions ($n_{\rm MPI}$) in PYTHIA8 is obtained by the master switch: PartonLevel:MPI = on. Technically, in hard interactions, $n_{\rm MPI}$ in PYTHIA8 is obtained as a ratio of integrated perturbative QCD 2 $\rightarrow$ 2 cross-sections (which depend on the minimum chosen value of $p_{T}$) and cross sections corresponding to the inelastic non-diffractive events \cite{Sjostrand:2017cdm}.

\begin{figure}
 \includegraphics[scale=0.45]{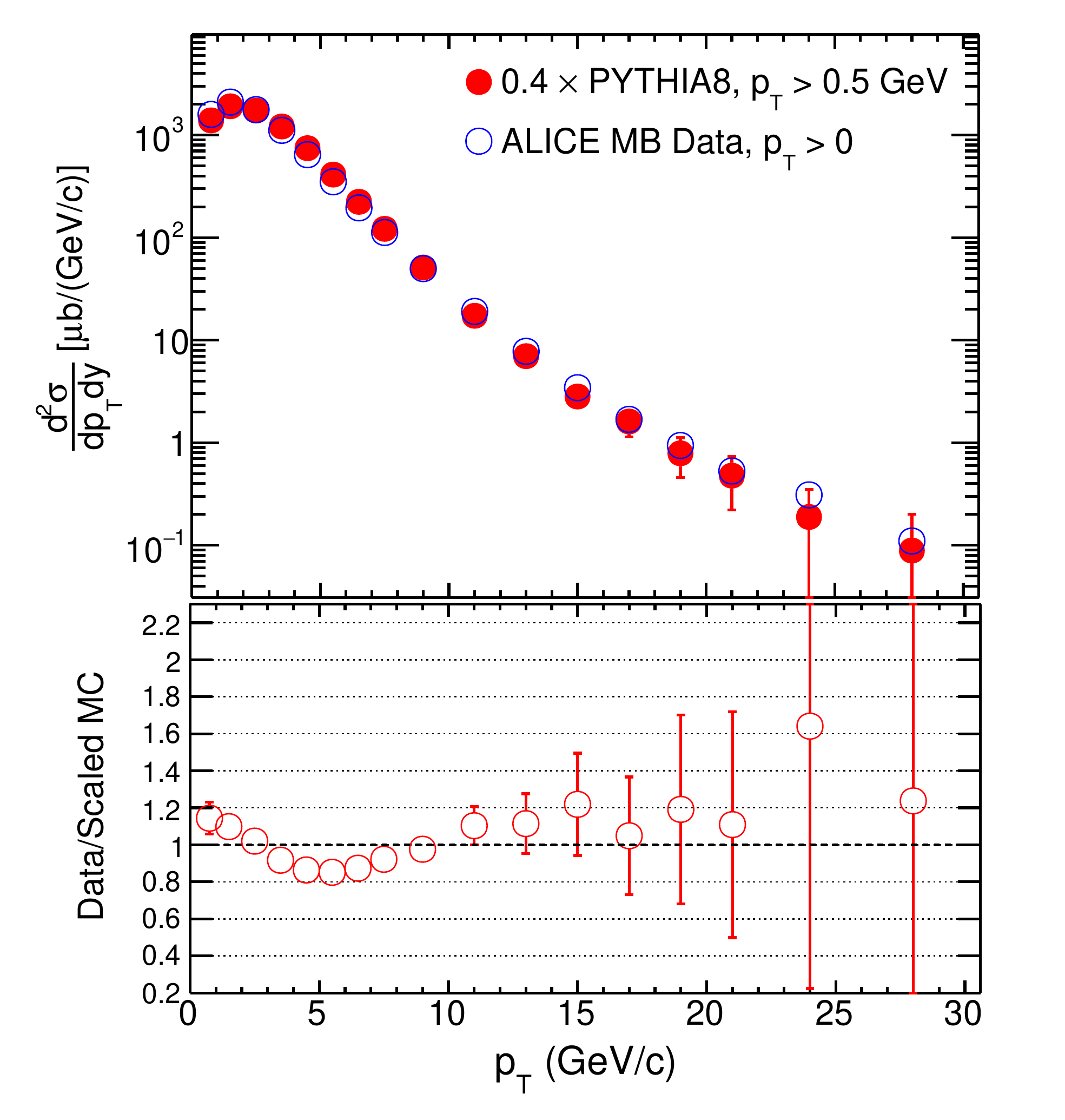}
\caption{(Color online) Upper panel shows the comparison of ALICE data~\cite{Acharya:2017hjh} and PYTHIA8 of J/$\psi$ production cross-section for $p+p$ collision in 2.5 $< y <$ 4.0  at $\sqrt{s}$ = 13 TeV. Lower panel shows the ratio between data and PYTHIA8.}
\label{fig:1}     
\end{figure}

\begin{figure}
  \includegraphics[scale=0.45]{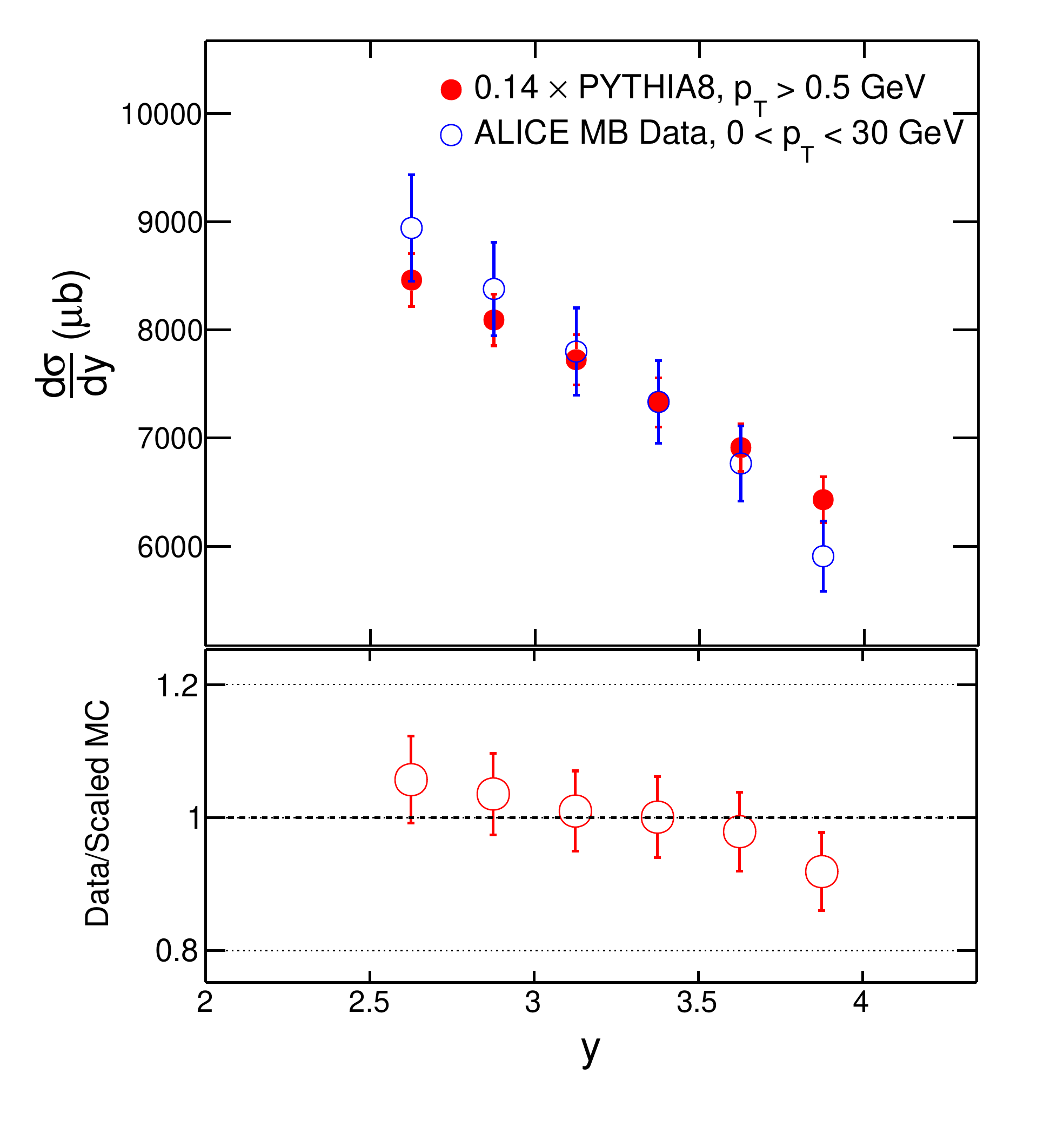}
\caption{(Color online) Upper panel shows the comparison of ALICE data~\cite{Acharya:2017hjh} and PYTHIA8 for d$\sigma$/dy of J/$\psi$ at $\sqrt{s}$ = 13 TeV. Lower panel shows the ratio between experimental data and PYTHIA8.}
\label{fig:2}     
\end{figure}

 We have generated 1600 million events for $p+p$ collisions at $\sqrt{s}$ = 13 TeV using PYTHIA8 for hard QCD processes and 100 million events each for 
$gg \rightarrow C$ and $q\bar{q} \rightarrow C$ processes, separetely. Study of J/$\psi$ production has been performed in the dimuon channel by forcing a J/$\psi$ to decay into dimuons ($\mu^{+}\mu^{-}$) in the MC simulation. The J/$\psi$ yield is then obtained through invariant mass reconstruction considering the detector acceptance. This helps in comparing the observations directly with the experimental data. 
The charged particle multiplicity yield, in particular the self-normalised yield, which is defined as $N_{\rm ch}^{i} / \langle  N_{\rm ch}  \rangle$, is calculated at mid-rapidity ($ |y| \textless 1.0$). Here $N_{\rm ch}^{i}$ is the mean of the charged particle multiplicity in a particular bin and $\langle  N_{\rm ch} \rangle$ is the mean of the charged particle multiplicity in minimum bias events. Here $i$ stands for multiplicity class. The relative J/$\psi$ yield is calculated in forward rapidity ($2.5 \textless   y \textless  4.0$) using the following relation:
\begin{equation}
\frac{Y_{\rm J/\psi}^{i}}{\langle  Y_{\rm J/\psi} \rangle}= \frac{N_{\rm J/\psi}^{i}}{N_{\rm J/\psi}^{\rm total}}\frac{N_{\rm evt}^{\rm total}}{N_{\rm evt}^{i}},
\label{eq1}
\end{equation}

where $N_{\rm J/\psi}^{i}$ and $N_{\rm evt}^{i}$ are the number of J/$\psi$ and number of events in $i^{th}$ multiplicity bin, respectively. 
$N_{\rm J/\psi}^{\rm total}$ and $N_{\rm evt}^{\rm total}$ are the total number of J/$\psi$ produced and total number of minimum bias events, respectively. As the frequency
of lower multiplicity events is higher, the bin width is taken smaller at lower multiplicity and then subsequently higher to maximize the statistics 
at high-multiplicity bins. 

The statistical uncertainties are calculated in each multiplicity bin for both relative charged particle multiplicity ($N_{\rm ch}^{i}/\langle  N_{\rm ch} \rangle$) and relative J/$\psi$ yield ($N_{\rm J/\psi}/ ~\langle  N_{\rm J/\psi}  \rangle$). Uncertainty in $N_{\rm ch}$ measurement is given by the ratio of RMS value of the charged particle multiplicity and square root of the number of charged particles in that bin ($N_{\rm RMS}^{\rm ch}/\sqrt{N_{\rm bin}^{\rm ch}}$). The ratio between RMS value of the minimum bias (MB) charged particle multiplicity and square root of the number of minimum bias charged particles ($N_{\rm RMS}^{\rm MB-ch}/\sqrt{N_{\rm MB}^{\rm ch}}$) gives the uncertainty in $\langle  N_{\rm ch} \rangle$. The uncertainty to measure the number of J/$\psi$ particles is simply $\sqrt{N_{\rm J/\psi}}$. These uncertainties are propagated using standard error propagation formula to estimate the uncertainties in relative charged particle multiplicity as well as in relative J/$\psi$ yield.

The mean transverse momentum, $\langle  p_{\rm T} \rangle$, of J/$\psi$ is calculated for each multiplicity bin and corresponding error is given by the ratio of the standard deviation (SD) of the $p_{\rm T}$ spectrum and square root of the number of entries in that multiplicity bin (${\rm SD}/\sqrt{N_{\rm mult-bin}^{p_{\rm T}}}$).

\section{Results and discussion}
 \label{result}
 To check the compatibility of PYTHIA8 with the experimental data, we have compared the ALICE production cross-section of J/$\psi$ with the transverse momentum distribution and rapidity distribution of PYTHIA8 simulated data in the same kinematic range.
 Fig.~\ref{fig:1} and Fig.~\ref{fig:2} show the comparison of J/$\psi$ production cross-section in $p+p$ collisions as a function of $p_{\rm T}$ and rapidity ($y$), respectively for minimum bias events. The open symbols represent the data obtained by ALICE experiment \cite{Acharya:2017hjh} and the solid circles show the results from PYTHIA8 event generator in $p+p$ collisions at $\sqrt{s}$ = 13 TeV. In order to see how well the spectral shapes of $p_{\rm T}$ and $y$  obtained from PYTHIA8 simulation match with the experimental data, we have used some arbitrary multipliers. The results obtained from PYTHIA8 are multiplied by a constant factor in each case (0.4 for $p_{\rm T}$ spectra and 0.14 for d$\sigma$/dy) to put them on the same footing as the ALICE experimental data. The used scaling factors are to check the matching of the spectral shapes and bear no physical significance. The lower panels of the figures show the ratio between experimental data and simulated data. It is observed that PYTHIA8 approximately reproduces the experimental data after scaling. For a low-$p_{\rm T}$ bin ($p_{\rm T} \sim 0.5$ GeV/c), the deviation is larger and for two high-$p_{\rm T}$ bins ($p_{\rm T} \sim 24$ GeV/c  and $p_{\rm T} \sim 28$ GeV/c), deviation is around 64\% and 23\%, respectively from the scaled MC. The deviations at higher $p_{\rm T}$ are more difficult to quantify due to lack of statistics. Furthermore, the rapidity spectra are well reproduced by scaled simulated data with a maximum (10-15)\% deviation from experimental data. This study provides us the confidence for further analysis of quarkonia production using PYTHIA8 in $p+p$  collisions at the LHC energies.
 
\subsection{\bf{Multiplicity dependence of J/$\psi$ production in hard-QCD processes}}
\label{hard_qcd}
\begin{figure}
  \includegraphics[scale=0.43]{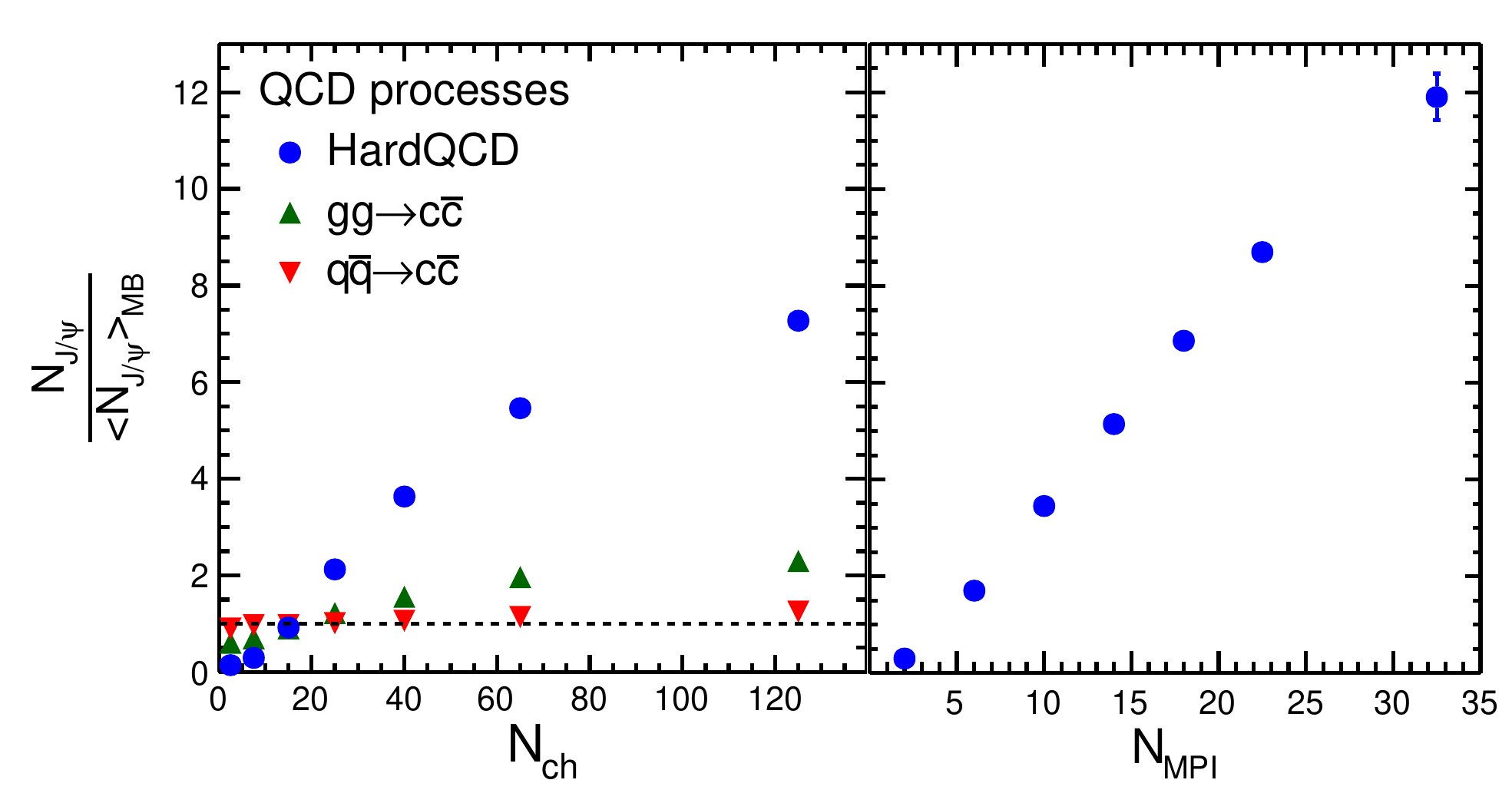}
 \caption  {(Color online) Left panel: Relative yield of J/$\psi$ as a function of charged particle multiplicity. Different symbols are for different hard processes of PYTHIA8.
 Right panel: Relative yield of J/$\psi$ as a function of number of multi-parton interactions for inclusive hard processes using PYTHIA8. The vertical lines in simulated data points are the 
 statistical uncertainties.}
 \label{fig:3}   
\end{figure}

From the first measurement of J/$\psi$ as a function of multiplicity, there have been constant efforts to understand the underlying physics. It is expected that the linear increase of J/$\psi$  production with event multiplicity is because of participation of higher number of gluons at higher multiplicities~\cite{Kopeliovich:2013yfa}. It is also believed that a major contribution to J/$\psi$ production comes from semi-hard interactions of partons in addition to first hardest one \cite{PorteboeufHoussais:2012gn}. In ref.~\cite{Thakur:2017kpv}, J/$\psi$ production with event multiplicity in $p+p$ 
collisions at $\sqrt{s}$= 7 TeV from experimental data~\cite{Abelev:2012rz} have been compared with PYTHIA8 simulated data. This comparison clearly shows that 4C-tuned PYTHIA8 {\it with} and {\it without} CR qualitatively explains the experimental data. To explore the effect of gluon and quark contributions on quarkonia production in $p+p$ collisions at $\sqrt{s}$ = 13 TeV, we have studied different processes of J/$\psi$ production in PYTHIA8  for different multiplicity classes. The left panel of Fig.~\ref{fig:3} shows the relative J/$\psi$ yield as a function of charged particle multiplicity for different processes such as, $gg \rightarrow C$, $q \bar{q} \rightarrow C$ and inclusive hard processes which contain all QCD processes. It is observed that all the processes are comparable up to $N_{\rm ch}$ $\sim$ 20-30 and for $N_{\rm ch} \ge  20-30 $ the contributions from $gg \rightarrow C$ and $q \bar{q} \rightarrow C$ become negligible compared to the inclusive hard processes. One of the main reasons behind this observation is the dominance of MPI at high-multiplicity events. The right panel of Fig.~\ref{fig:3} supports this statement as we can see the relative J/$\psi$ yield to increase linearly with number of MPIs. The relative (with respect to minimum bias collisions) yield of J/$\psi$ is more for $gg \rightarrow C$ compared to $q \bar{q} \rightarrow C$ processes. This might be due to the presence of large gluon densities in the parton distribution function of the colliding protons \cite{Kopeliovich:2013yfa}.
 
\begin{figure}
  \includegraphics[scale=0.42]{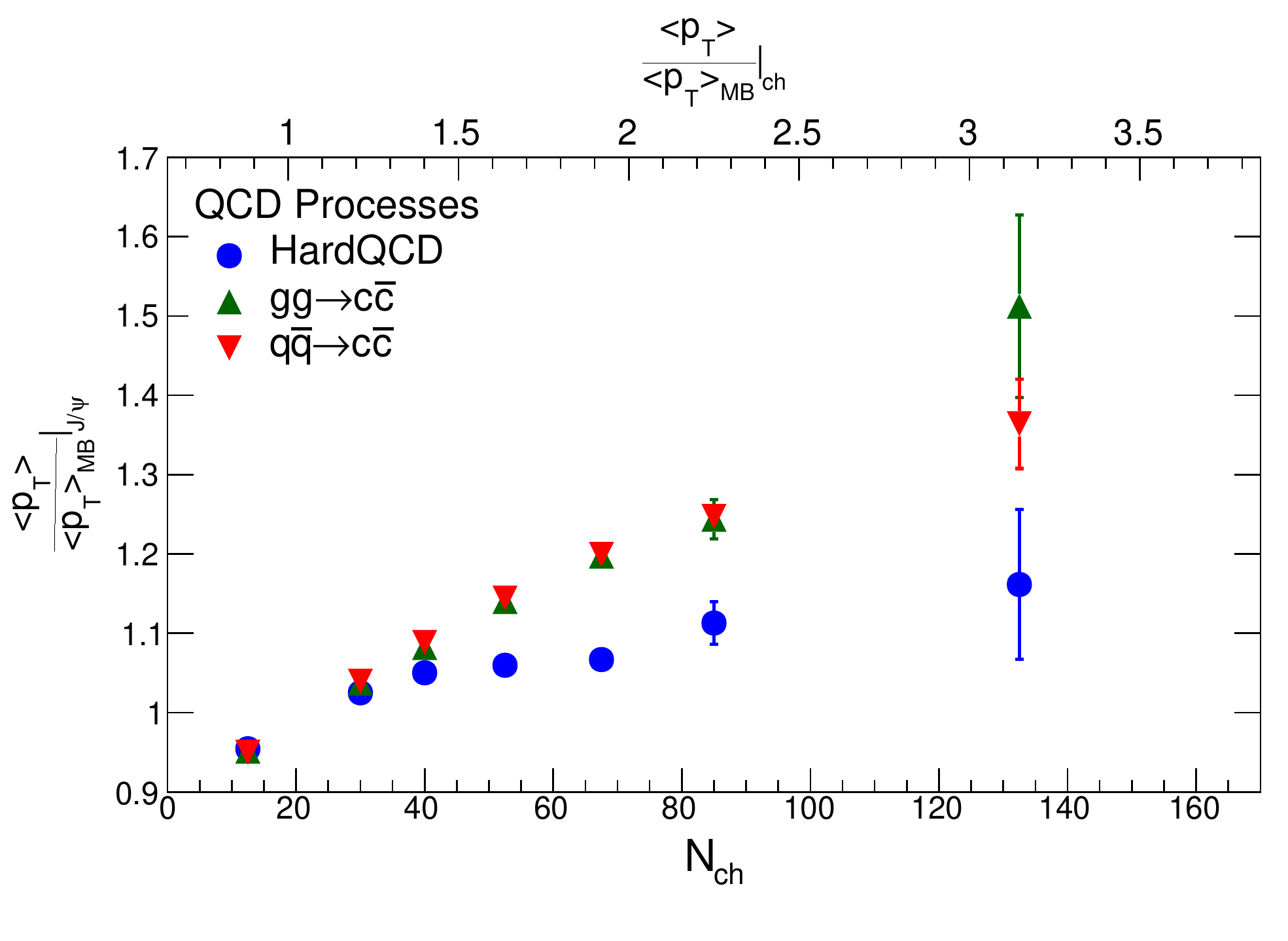}
 \caption  {(Color online) Relative mean transverse momentum of J/$\psi$ is presented as a function of charged particle multiplicity (in lower $x$-axis) as well as with relative $\langle  p_{\rm T} \rangle$ of charged particles (upper $x$-axis) for different hard processes of PYTHIA8. Different symbols are for different processes. The vertical lines in simulated data points are the statistical uncertainties.} 
\label{fig:4}
\end{figure}

Figure~\ref{fig:4} shows relative mean transverse momentum of J/$\psi$ for different QCD processes of PYTHIA8 as a function of charged particle multiplicity {\bf ($N_{\rm ch}$)}. It is observed that relative $\langle  p_{\rm T} \rangle$ increases with multiplicity. It indicates that harder J/$\psi$ are produced in higher multiplicity classes. However, for $gg \rightarrow C$ and $q \bar{q} \rightarrow C,$ the rate of increase is much faster than that of inclusive hard processes. It is found that up to $N_{\rm ch} \sim$  (20-30) the contributions to $\langle  p_{\rm T} \rangle$ mainly come from the processes: $gg \rightarrow C$ and $q \bar{q} \rightarrow C$, and for $N_{\rm ch} \ge$ (20-30) the rate of increase of $\langle  p_{\rm T} \rangle$ is slower for inclusive hard processes than that of other two processes. This may be due to the dominance of MPIs (contribution of semi-hard J/$\psi$ from MPI), as the second interaction will not be as hard as the first one. This is because of the inclusion of  rescattering effects in PYTHIA8~\cite{Corke:2009tk}. Rescattering in PYTHIA is defined as one where one parton may undergo successive collisions against several other partons. Infact, the J/$\psi$ yield and $\langle  p_{\rm T} \rangle$ versus multiplicity are the complementary studies. The yield versus multiplicity tells about the role of 
MPI in J/$\psi$ production, whereas $\langle  p_{\rm T} \rangle$ versus multiplicity gives information about the time ordering of MPI in J/$\psi$ production. The J/$\psi$ produced from the later stage of MPIs reduce the value of $\langle  p_{\rm T} \rangle$.  Higher $p_{\rm{T}}$ refers to earlier time: giving a natural time-ordering. In addition to the rescattering effects, $\langle  p_{\rm T} \rangle$ can have contribution from color reconnection where the lower $p_{\rm T}$ partons are merged with the ones in higher $p_{\rm T}$. Within uncertainties, the $\langle  p_{\rm T} \rangle$ of J/$\psi$ for $gg \rightarrow C$ and $q \bar{q} \rightarrow C$ is observed to be nearly the same. This indicates that the contributions from gluon fusion and light-quark annihilation to J/$\psi$ production increase in equal proportion with multiplicity. 
The relative $\langle  p_{\rm T} \rangle$ of the J/$\psi$ in inclusive hard processes shows a slightly increasing behaviour as a function of $N_{\rm ch}$ unlike that in p-Pb collisions as observed by 
ALICE experiment \cite{Adamova:2017uhu}. 
 
For completeness and for a clear view of the observed pictures, it is worth investigating the relative hardness versus relative softness. Although, there is no such type of experimental study available, we have tried to study this using PYTHIA8. Figure~\ref{fig:4} represents the relative $\langle  p_{\rm T} \rangle$ of J/$\psi$ (hardness) as a function of relative $\langle  p_{\rm T} \rangle$ of charged particles (softness) (in upper $x$-axis). It can be clearly seen that the processes: $gg \rightarrow C$ and  $q \bar{q} \rightarrow C$ give an approximately linear growth in $\langle  p_{\rm T} \rangle$ while the ``all hard QCD" results are slower than linear. Further, the current study shows that, the $\langle  p_{\rm T} \rangle$ of charged particles increases much faster compared to $\langle  p_{\rm T} \rangle$ of J/$\psi$. The reason behind this difference is the rescattering effect which is implemented in PYTHIA8. For example, a pion can cause a larger $p_{\rm T}$ increase of another pion than that of much more massive J/$\psi$. Thus, in a system of many particles, the light hadrons could interact multiple times which results in strongly increasing $p_{\rm T}$ with increasing multiplicity.


\subsection{\bf{Transverse momentum and multiplicity dependence of J/$\psi$ production}}
 \label{pt_mult}
\begin{figure}.
   \includegraphics[scale=0.425]{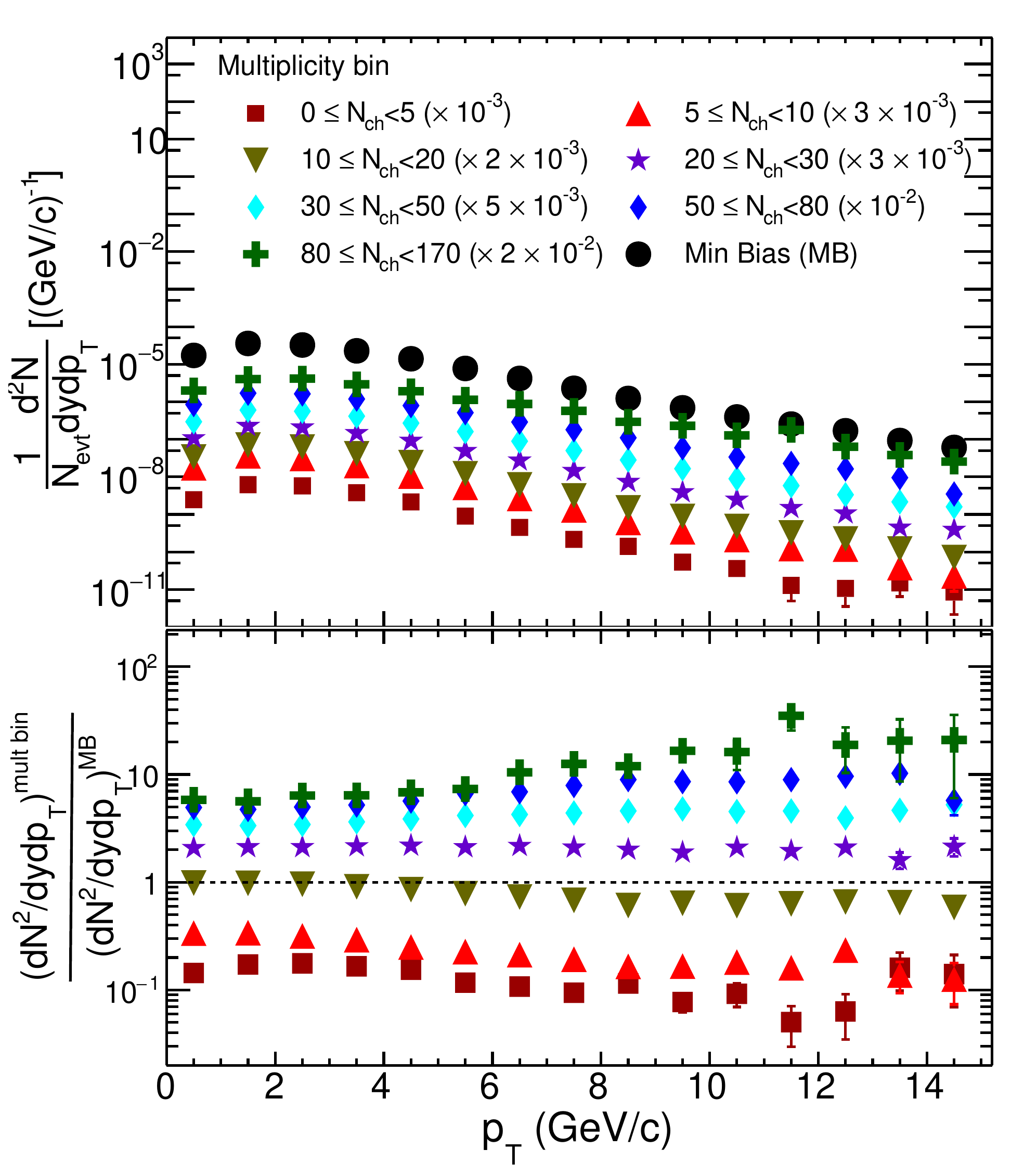}
\caption {(Color online) Upper panel: Transverse momentum spectra of J/$\psi$ for different multiplicity classes. Different symbols represent different multiplicity classes. Solid black circles show the spectrum for minimum bias events. Lower panel: Ratio of the spectra of different multiplicity classes with respect to the spectra of MB events without the scale factor. The error bars in the simulated data points are the statistical uncertainties.}
 \label{fig6} 
\end{figure}

Figure \ref{fig6}  shows the transverse momentum spectra of J/$\psi$ for (0 $\leq$ $N_{\rm ch}< $ 5), (5 $\leq$ $N_{\rm ch} < $ 10), (10 $\leq$ $N_{\rm ch} < $ 20), (20 $\leq$ $N_{\rm ch} < $ 30), 
(30 $\leq$ $N_{\rm ch} < $ 50), (50 $\leq$ $N_{\rm ch} < $ 80), (80 $\leq$ $N_{\rm ch} < $ 170) multiplicity classes along with the minimum bias (MB) events. The spectra are multiplied by constant factors to get clearer view of each spectrum. Lower panel of the figure shows the ratio of the spectra (without scale factor) of different multiplicity classes with respect to the spectra of MB events. The uncertainties on the points are statistical. We have observed that for high-multiplicity bins, the number of J/$\psi$ increases with increase of $p_{\rm T}$. This indicates that as we
go from low to high multiplicities, harder J/$\psi$ are produced. The ratio in the lower panel of Fig. ~\ref{fig6}  is observed to be divided into two parts around $N_{\rm ch}(10-20)$: less than 1 and greater than 1. This tells that MPIs start to dominate for the production of J/$\psi$ from $N_{\rm ch } > $ 20. This observation is not only in line with our earlier work \cite{Thakur:2017kpv} but also consistent with Figs.~\ref{fig:3} and~\ref{fig:4}.

\begin{figure}
    \includegraphics[scale=0.42]{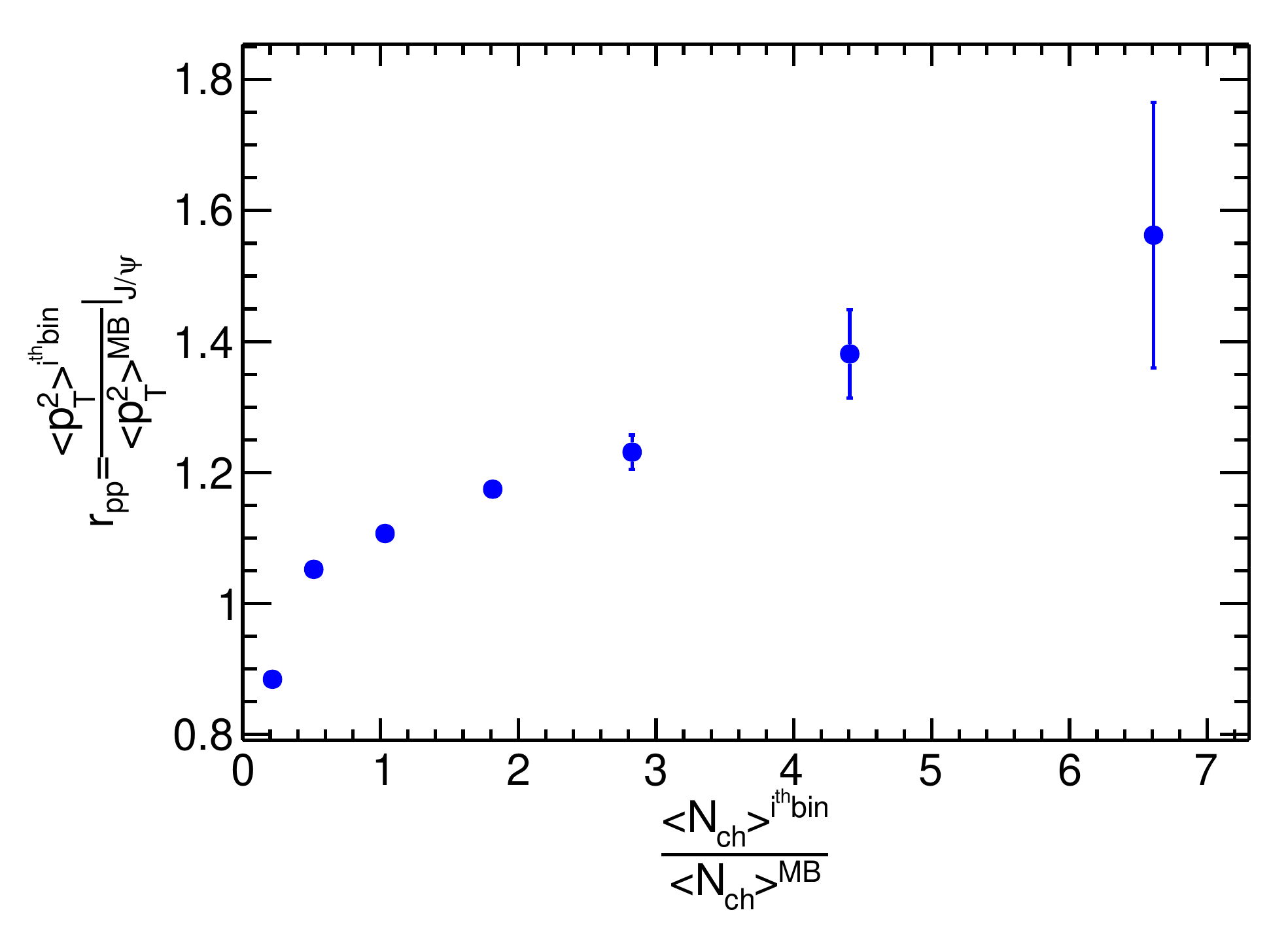}
 \caption{(Color online) The newly defined nuclear modification factor $r_{\rm pp}$ for J/$\psi$ as a function of relative charged multiplicity. The error bars show the statistical uncertainties.}
 \label{fig7}
\end{figure}

\subsection{\bf{Multiplicity dependence of $r_{pp}$}}
\label{rpp}

To understand the J/$\psi$ production mechanisms in high-multiplicity events in $p+p$ collisions at $\sqrt{s}$ = 13 TeV, we study a newly defined nuclear 
modification factor using transverse momentum \cite{Zhou:2014kka,Zhou:2009vz}:
\begin{equation}
r_{\rm pp} = \frac{\langle p_{\rm T}^{2} \rangle_{i}}{\langle p_{\rm T}^{2} \rangle_{\rm MB}},
\label{eq2}
\end{equation}
where $\langle p_{\rm T}^{2} \rangle_{i}$ and $\langle p_{\rm T}^{2} \rangle_{\rm MB}$ are the averaged transverse momentum squared for J/$\psi$ in $i^{th}$ 
multiplicity class and minimum biased events, respectively. $\langle p_{\rm T}^{2}\rangle$ is generally attributed to the multi-scattering of partons in
the initial state. This can be treated as a random walk in transverse momentum space and the observed $\langle p_{\rm T}^{2}\rangle$ is predicted to increase linearly with the mean path length of the traversed parton. Therefore, $r_{\rm pp}$ can be used as a good observable to study the difference of high-multiplicity $p+p$ events with respect to minimum bias events. Figure~\ref{fig7} shows the $r_{\rm pp}$ of J/$\psi$ as a function of relative charged particle multiplicity using PYTHIA. It is observed that $r_{\rm pp}$ increases with increasing multiplicity. The increasing trend of $r_{\rm pp}$ reveals larger system size at high-multiplicity compared to low multiplicity events. To understand the results, we need to compare it with the results obtained in heavy-ion collisions. Reference~\cite{Zhou:2014kka} shows that the $r_{\rm AA}$ (which is defined as $(\langle p_{\rm T}^{2} \rangle_{\rm AA})/(\langle p_{\rm T}^{2} \rangle_{\rm pp})$ values are different for different collision energies in heavy-ion collisions. It is found that $r_{\rm AA}  >$ 1 for SPS, 
$r_{\rm AA} \sim$ 1 for RHIC and $r_{\rm AA} < $ 1 for LHC in mid-rapidity region. This is because of the fact that at SPS energies almost all the measured J/$\psi$ are produced via initial hard processes and the increase with centrality arises from the Cronin effect and the leakage effect~\cite{Zhu:2004nw}. But at RHIC and LHC, regeneration plays an important role. At RHIC, regeneration and initial hard processes cancel each other and we get $r_{\rm AA} \sim$ 1. Whereas at LHC, regeneration dominates and we get decreasing $r_{\rm AA}$ with centrality. We observe that for $p+p$ collisions, $r_{\rm pp}$ values show a trend similar to SPS heavy-ion results. This indicates that even at high centre-of-mass energy and for highest multiplicity classes in $p+p$ collisions, regeneration effect is negligible and initial hard processes dominate in J/$\psi$ production. 


\begin{figure}[hbt!]
\includegraphics[width=4.16 cm, height=6.5 cm]{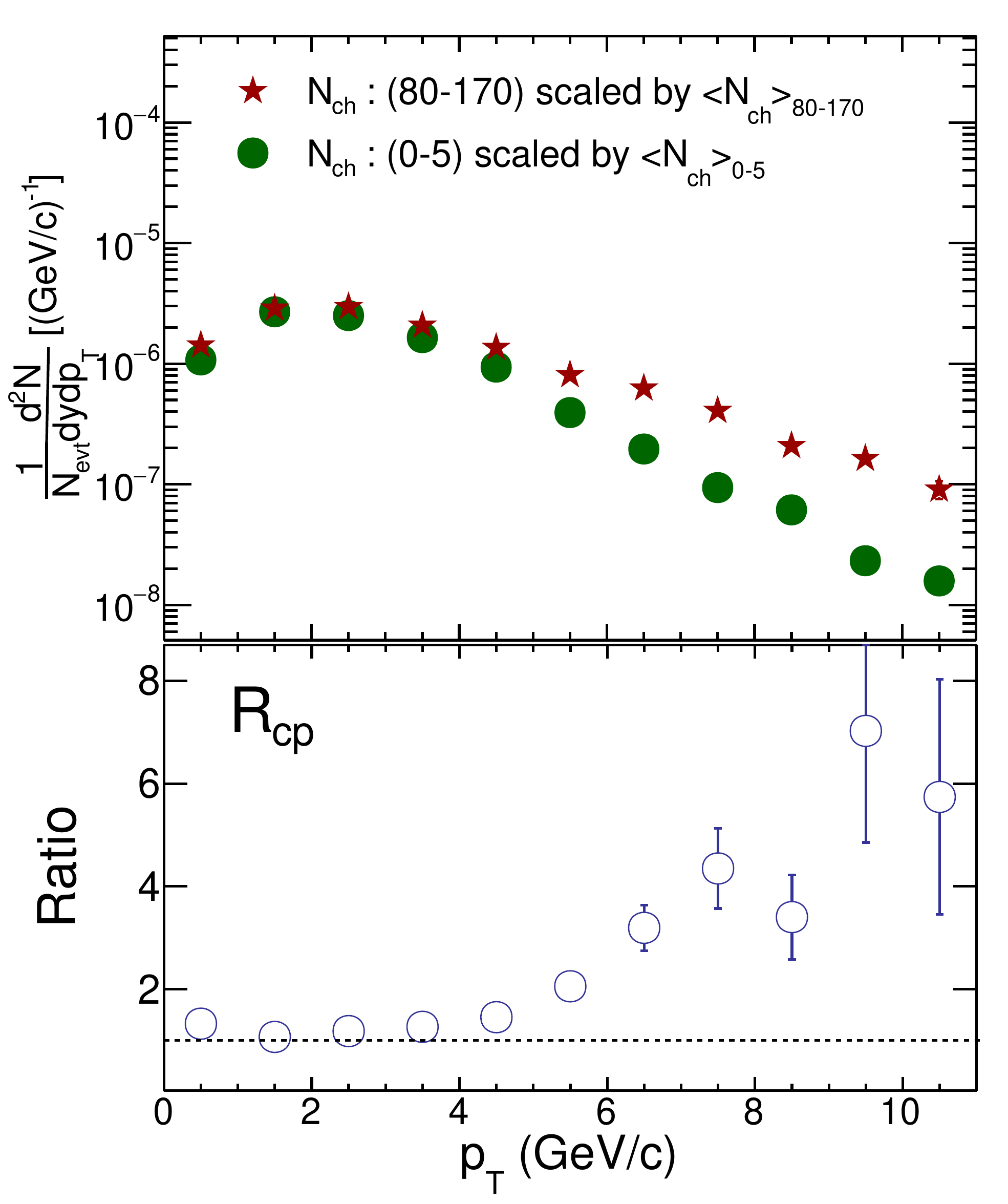}
 \includegraphics[width=4.16 cm, height=6.5 cm]{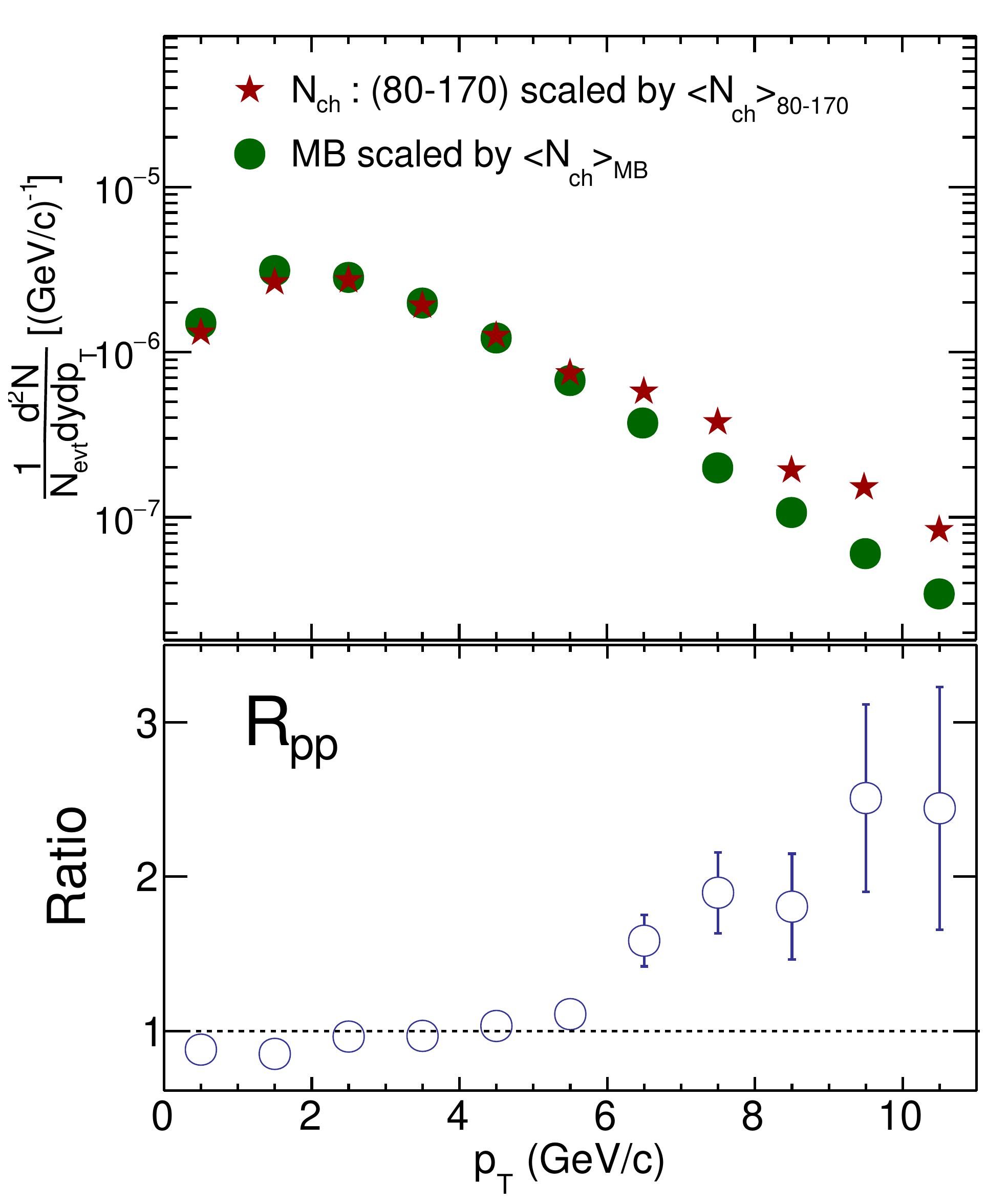}
 \caption{(Color online) Left upper panel shows the invariant yield as a function of $p_{\rm T}$ for different multiplicity classes and are scaled by corresponding $\langle N_{\rm ch} \rangle$. Stars are for highest multiplicity bin: (80-170) and circles represent lowest multiplicity bin: (0-5). Left lower panel shows $R_{\rm cp}$ as a function of $p_{\rm T}$. Right upper panel shows the invariant yield as a function of $p_{\rm T}$ for high multiplicity bin and MB events, and are scaled by corresponding $\langle N_{\rm ch} \rangle$. Stars are for highest multiplicity bin: (80-170) and circles represent MB 
events. Right lower panel shows $R_{\rm pp}$ as a function of $p_{\rm T}$.}
 \label{fig8}
\end{figure}

\subsection{\bf{Ratio of particle production yields}}
 \label{R_pp}
 Charmonium suppression is a universally accepted probe for the deconfined medium in heavy-ion collisions. To understand the suppression, it is necessary to understand J/$\psi$ production in $p+p$ collisions as well as related effects which arise not because of system formation rather due to other kinds of underlying effects on the observables. For example, in order to understand the possibility of formation of a system in high-multiplicity events in $p+p$ collisions at $\sqrt{s}$ = 13 TeV, we define two variables based on Ref.~\cite{Zhang:2018hcp} as:
 
 \begin{equation}
 R_{\rm pp}=\frac{\langle  N_{\rm ch} \rangle_{\rm MB}}{\langle  N_{\rm ch} \rangle_{80-170}}\frac{[(1/N_{\rm evt})(dN/dp_{\rm T})]_{80-170}}{[(1/N_{\rm evt})(dN/dp_{\rm T})]_{\rm MB}},
 \label{eq3}
\end{equation}
\begin{equation}
R_{\rm cp}=\frac{\langle  N_{\rm ch} \rangle_{0-5}}{\langle  N_{\rm ch} \rangle_{80-170}}\frac{[(1/N_{\rm evt})(dN/dp_{\rm T})]_{80-170}}{[(1/N_{\rm evt})(dN/dp_{\rm T})]_{0-5}},
 \label{eq4}
 \end{equation}
 
which are similar to the nuclear modification factors $R_{\rm AA}$ and $R_{\rm cp}$ in heavy-ion collisions. Here, $\langle  N_{\rm ch} \rangle_{80-170}$, $ \langle  N_{\rm ch} \rangle_{0-5}$ and $\langle  N_{\rm ch} \rangle_{\rm MB}$ are the average charged particle multiplicity for ($80 \leq N_{\rm ch} < 170$), ($0 \leq N_{\rm ch} <  5 $) multiplicity classes and MB events, respectively. Since MPIs are proportional to $N_{\rm ch}$ we take $\langle  N_{\rm ch} \rangle$ as the scaling factor to measure the $R_{\rm pp}$ and $R_{\rm cp}$ for $p+p$ collisions. Figure~\ref{fig8} shows $R_{\rm pp}$ and $R_{\rm cp}$ as a function of $p_{\rm T}$ calculated using PYTHIA8. Right upper panel of  Fig.~\ref{fig8} shows the $p_{\rm T}$-spectra of J$/\psi$ for highest (80-170) multiplicity class and MB events scaled with corresponding $\langle  N_{\rm ch} \rangle$ values. Lower panel shows the $R_{\rm pp}$ as defined in Eq.~\ref{eq3}. Around 10\% suppression is observed for $p_{\rm T} < $ 2 GeV/c. 
For mid-$p_{\rm T}$ region: 2-6 GeV/c, $R_{\rm pp}$ values are unity and increase rapidly in the high-$p_{\rm T}$ region {\it i.e.} $p_{\rm T} >$ 6 GeV/c. $R_{\rm pp}$ (or $R_{\rm AA}$) indicates that J/$\psi$ particles lose their momentum in the QCD medium formed in high-multiplicity $p+p$ collisions, thereby resulting in the spectrum to shift towards lower-$p_{\rm T}$ (becomes steeper). Since there is a hint of suppression at low-$p_{\rm T}$ region, one can draw a conclusion that the QCD medium formed in high-multiplicity events is different from that of minimum bias events, which needs further investigation. PYTHIA does not incorporate system formation, still it shows little suppression, therefore, it will be interesting to measure $R_{\rm pp}$ in experiments, so that, one can conclude about the observables like $R_{\rm AA}$ (or $R_{\rm pp}$). This study reveals that to understand the formation of Quark Gluon Plasma through J/$\psi$ suppression in heavy-ion collisions, it is certainly necessary to understand the similarity/difference between the QCD medium formed in $p+p$ and heavy-ion collisions. The $R_{\rm cp}$ is consistent with unity up to $p_{\rm T} <$ 4 GeV/c and increases at high-$p_{\rm T}$ similar to $R_{\rm pp}$.  

\section{Summary}
  \label{sum}
 In summary, multiplicity dependent study of the transverse momentum spectra of J/$\psi$  has been performed using 4C tuned PYTHIA8 
 MC event generator in $p+p$ collisions at $\sqrt{s}$ = 13 TeV at forward rapidity ($2.5 <   y <  4.0$). In this work, we have tried to understand the J/$\psi$ production mechanism as a function of multiplicity as well as $p_{\rm T}$. Relative J/$\psi$ yields and $\langle p_{\rm T} \rangle$ are measured as a function of charged particle multiplicity for different hard-QCD processes. Also, relative $\langle  p_{\rm T} \rangle$ is measured as a function of charged particle multiplicity for different hard processes. The ratio of $\langle  p_{\rm T}^{2} \rangle$ between $i^{th}$ multiplicity class and minimum biased events, which is defined as $r_{\rm pp}$, is measured as a function of relative charged particle multiplicity. For the first time, we have proposed and simulated $R_{\rm pp}$ and $R_{\rm cp}$ in $p+p$ collisions. The conclusions of our study on J/$\psi$ using PYTHIA8 4C production are the following:
 
  \begin{itemize}
  \item[$\bullet$] Up to $N_{\rm ch}\simeq$  (20-30), J/$\psi$ are mainly produced via $gg \rightarrow C$ and 
 $q \bar{q} \rightarrow C$, and for $N_{\rm ch} \ge$ (20-30) the contribution from MPI dominates. This leads to a decrease in $\langle  p_{\rm T} \rangle$ of inclusive J/$\psi$ when studied with respect to charged particle multiplicity in MB collisions due to contribution of J/$\psi$ from semi-hard processes.

 \item[$\bullet$] Dominance of J/$\psi$ production for inclusive hard-QCD processes beyond $N_{\rm ch} \ge$ (20-30) compared to $gg \rightarrow C$ and $q \bar{q} \rightarrow C$ processes indicates that MPI plays an important role for events with $N_{\rm ch} \ge$ (20-30).

 \item[$\bullet$ ] It is found that relative $\langle  p_{\rm T} \rangle$ of J/$\psi$ for $gg \rightarrow C$ and $q \bar{q} \rightarrow C$  processes remain almost the same within uncertainties for PYTHIA. 
  
 \item[$\bullet$] From the $p_{\rm T}$-spectra of J/$\psi$ for different multiplicity classes, we found that harder J/$\psi$ are produced as we go towards the higher multiplicities.
 
 \item[$\bullet$] A new observable, $r_{\rm pp}$ is introduced to study possible medium effects in high-multiplicity $p+p$ collisions. The increasing trend of $r_{\rm pp}$ reveals larger system size at high-multiplicity compared to low multiplicity events. 
 
 
 \item[$\bullet$] $R_{\rm pp}$ shows around 10\% suppression for $p_{\rm T} < $ 2 GeV/c. However, there is no suppression  observed in case of $R_{\rm cp}$
 measurements. This indicates that the QCD medium formed in high-multiplicity $p+p$ collisions is different from that of minimum bias events. However, experimental measurements of these quantities are necessary to have solid conclusions.
 
 \end{itemize}
 
 \section{Acknowledgement} 
DT acknowledges UGC, New Delhi, Government of India for financial supports.  SD and RNS acknowledge the financial supports  from  ALICE  Project  No. SR/MF/PS-01/2014-IITI(G) of  
Department  of  Science $\&$ Technology,  Government of India. The authors gratefully acknowledge Professor Leif L\"onnblad for valuable discussions. This research used resources of 
the LHC grid computing facility at Variable Energy Cyclotron Center, Kolkata. We would like to thank Prof. B.K. Nandi for carefully reading the final version of the manuscript and providing useful comments.

\appendix
\setcounter{secnumdepth}{0}
\section{Appendix}
For completeness, we have listed the numerical values of relative yield (Table 1) and relative $\langle  p_{\rm T} \rangle$ (Table 2) of J/$\psi$ for HardQCD, gg$\rightarrow C$ and q$\bar{q}\rightarrow C$ processes as a  function of multiplicity ($N_{\rm ch}$-bin) along with their uncertainties.

\begin{table*}[h] \small
\centering
\caption{Relative yield of  J/$\psi$ {\bf $(N_{\rm J/\psi}^{i}/N_{\rm J/\psi}^{\rm MB})$} for hardQCD, gg$\rightarrow C$, q$\bar{q}\rightarrow C$ processes as a  function of charged particle multiplicity in $p+p$ collision at $\sqrt{s}$ = 13 TeV using PYTHIA.}
\small
\begin{center}
\begin{tabular}{|c|c|c|c|c|c|}
\hline
 $N_{\rm ch}$-bin&   	HardQCD         		   &  gg$\rightarrow C$    	   & q$\bar{q}\rightarrow C$     \\ \hline
   0-5    	  &0.1411 $\pm$ 0.0008                      &0.6041 $\pm$ 0.0009                          &0.8979 $\pm$0.0008     \\ \hline
  5-10    	&0.2922 $\pm$ 0.0008             	 	 & 0.6972$\pm$ 0.0007                            &0.9718 $\pm$ 0.0006           \\ \hline
 10-20    	& 0.9211$\pm$ 0.0018             		 &0.9025$\pm$ 0.0007                             & 0.9690$\pm$0.0005 \\ \hline
 20-30    	&2.1236$\pm$ 0.0042                     	&1.2188$\pm$ 0.0012                               & 1.0094 $\pm$ 0.0006      \\ \hline
 30-50  	 &3.6313 $\pm$ 0.0065                     	 &1.5571$\pm$ 0.0015                               &1.0600$\pm$ 0.0006    \\ \hline
 50-80  	 &5.4608 $\pm$ 0.0162                    	&1.9631 $\pm$ 0.0034                                &1.1361 $\pm$ 0.0011         \\ \hline
 80-170  	&7.2709$\pm$ 0.0954                 	&2.2976 $\pm$  0.0203                               & 1.2470$\pm$ 0.0060                 \\ \hline
\end{tabular}
\end{center}
\label{tablefin}
\label{table1}
\end{table*}


\begin{table*}[h]\small
\caption{Relative $\langle  p_{\rm T} \rangle$ of J/$\psi$ $(\langle  p_{\rm T} \rangle_{\rm J/\psi}^{i}/\langle  p_{\rm T} \rangle_{\rm J/\psi}^{\rm MB})$ as a function of charged particle multiplicity for hardQCD, gg$\rightarrow C$, q$\bar{q}\rightarrow C$ processes in $p+p$ collision at $\sqrt{s}$ = 13 TeV using PYTHIA.}
\small
\begin{center}
\begin{tabular}{|c|c|c|c|c|c|}
\hline
 $N_{\rm ch}$-bin&  	 HardQCD            		&  gg$\rightarrow C$    		& q$\bar{q}\rightarrow C$    \\ \hline
   0-25     		 &0.9545$\pm$ 0.0033       	 &0.9516 $\pm$ 0.0017         		&0.9499 $\pm$0.0069   			       \\ \hline
  25-35   		 &1.0250 $\pm$ 0.0049       	&1.0374 $\pm$ 0.0031           	 	&1.0396 $\pm$ 0.0016         	       \\ \hline
  35-45   		&1.0502 $\pm$ 0.0059       	&1.0821$\pm$ 0.0040           		 &1.0885 $\pm$ 0.0021  		       \\ \hline
  45-60   		&1.0597$\pm$ 0.0068          	&1.1401 $\pm$ 0.0052             	         &1.1443 $\pm$ 0.0027                   \\ \hline
  60-75   		&1.0669 $\pm$ 0.0124        	&1.1973$\pm$ 0.0105          		 &1.1994$\pm$ 0.0055                     \\ \hline
  75-95   		&1.1129 $\pm$ 0.0267        	&1.2439 $\pm$ 0.0246          		 &1.2472 $\pm$ 0.0129                        \\ \hline
  95-170 		 &1.1615 $\pm$ 0.0945      	 &1.5124 $\pm$  0.1150                     &1.1364 $\pm$ 0.0565                          \\ \hline
\end{tabular}
\end{center}
\label{tablefin}
\label{table1}
\end{table*}

 
 
 \end{document}